\documentclass[12pt]{article}
\usepackage{latexsym,amsmath}
\input amssym
\begin{document}
\renewcommand{\theequation}{\thesection.\arabic{equation}}
\def\prg#1{\medskip{\bf #1}}
\def\lra{\leftrightarrow}        \def\Ra{\Rightarrow}
\def\nin{\noindent}              \def\pd{\partial}
\def\dis{\displaystyle}          \def\dfrac{\dis\frac}
\def\grl{{GR$_\Lambda$}}         \def\vsm{\vspace{-10pt}}
\def\Lra{{\Leftrightarrow}}      \def\ads3{AdS$_3$}
\def\cs{{\scriptscriptstyle\rm CS}}
\def\Leff{\hbox{$\mit\L_{\hspace{.6pt}\rm eff}\,$}}
\def\Tr{\hbox{\rm Tr\hspace{1pt}}}
\def\gp{($\Bbb{P}$)}

\def\D{{\Delta}}      \def\bC{{\bar C}}     \def\bT{{\bar T}}
\def\bH{{\bar H}}     \def\bL{{\bar L}}     \def\bI{{\bar I}}
\def\hO{{\hat O}}     \def\hG{{\hat G}}     \def\tG{{\tilde G}}
\def\cL{{\cal L}}     \def\cM{{\cal M }}    \def\cE{{\cal E}}
\def\cA{{\cal A}}     \def\cI{{\cal I}}     \def\cC{{\cal C}}
\def\cF{{\cal F}}     \def\hcF{\hat{\cF}}   \def\bcF{{\bar\cF}}
\def\cH{{\cal H}}     \def\hcH{\hat{\cH}}   \def\bcH{{\bar\cH}}
\def\cK{{\cal K}}     \def\hcK{\hat{\cK}}   \def\bcK{{\bar\cK}}
\def\cO{{\cal O}}     \def\hcO{\hat{\cal O}} \def\tR{{\tilde R}}

\def\G{{\mit\Gamma}}  \def\S{\Sigma}        \def\L{{\mit\Lambda}}
\def\a{\alpha}        \def\b{\beta}         \def\g{\gamma}
\def\d{\delta}        \def\m{\mu}           \def\n{\nu}
\def\th{\theta}       \def\k{\kappa}        \def\l{\lambda}
\def\vphi{\varphi}    \def\ve{\varepsilon}  \def\p{\pi}
\def\r{\rho}          \def\Om{\Omega}       \def\om{\omega}
\def\s{\sigma}        \def\t{\tau}          \def\eps{\epsilon}
\def\ups{\upsilon}    \def\tom{{\tilde\om}} \def\bw{{\bar w}}
\def\nn{\nonumber}
\def\be{\begin{equation}}             \def\ee{\end{equation}}
\def\ba#1{\begin{array}{#1}}          \def\ea{\end{array}}
\def\bea{\begin{eqnarray} }           \def\eea{\end{eqnarray} }
\def\beann{\begin{eqnarray*} }        \def\eeann{\end{eqnarray*} }
\def\beal{\begin{eqalign}}            \def\eeal{\end{eqalign}}
\def\lab#1{\label{eq:#1}}             \def\eq#1{(\ref{eq:#1})}
\def\bsubeq{\begin{subequations}}     \def\esubeq{\end{subequations}}
\def\bitem{\begin{itemize}}           \def\eitem{\end{itemize}}

\title{Supersymmetric 3D gravity with torsion: asymptotic symmetries}

\author{B. Cvetkovi\'c and M. Blagojevi\'c
\footnote{Email addresses: {\tt cbranislav@phy.bg.ac.yu,
                                mb@phy.bg.ac.yu}} \\
Institute of Physics, P. O. Box 57, 11001 Belgrade, Serbia}
\date{}
\maketitle
\begin{abstract}
We study the structure of asymptotic symmetries in $N=1+1$
supersymmetric extension of three-dimensional gravity with torsion.
Using a natural generalization of the bosonic anti-de Sitter
asymptotic conditions, we show that the asymptotic Poisson bracket
algebra of the canonical generators has the form of two independent
super-Virasoro algebras with different central charges.
\end{abstract}

\section{Introduction}
\setcounter{equation}{0}

Contemporary research in the area of three-dimensional (3D) gravity is
motivated by enormous difficulties in our attempts to properly
understand fundamental features of both classical and quantum gravity.
In the traditional approach, based on general relativity (GR), the
gravitational dynamics has been investigated in the realm of {\it
Riemannian\/} geometry, leading to a number of outstanding results
\cite{1,2,3,4,5,6}. However, there is a more general, gauge-theoretic
conception of gravity based on {\it Riemann-Cartan\/} geometry, in
which both the curvature and the torsion characterize the geometric
structure of gravity \cite{7,8}. Within this, more general geometric
setting, one has a real chance to explore the influence of geometry on
the gravitational dynamics.

The application of these ideas to 3D gravity started in the early
1990s, when Mielke and Baeckler proposed a general topological action
for 3D gravity with torsion \cite{9}. Recent developments confirmed
that this approach offers a rich and respectable dynamical content.
In particular, it has been shown that: (i) 3D gravity with torsion
possesses the generalized BTZ black hole solution \cite{10,11} with
interesting thermodynamic properties \cite{12}, (ii) it can be
formulated as a Chern-Simons gauge theory \cite{11,13}, and (iii) with
suitably chosen asymptotic conditions, the asymptotic symmetry is
described by two independent Virasoro algebras with different central
charges \cite{14}.

The supersymmetric extension of 3D gravity with torsion, constructed
recently in \cite{15}, offers a new look into the underlying
geometric structure. In the present paper, we use the canonical
approach to analyze the asymptotic symmetry of $N=1+1$ supersymmetric
extension of this theory, which yields a deeper insight into the
dynamical role of torsion. Our investigation is intended to
generalize the results obtained in the anti-de Sitter (AdS) sector of
Riemannian GR \cite{16} (see also \cite{17,18}), to 3D gravity with
torsion. More precisely, we will demonstrate that a suitable
extension of the AdS asymptotics to the fermionic sector produces the
super-conformal symmetry in the asymptotic region.

The paper is organized as follows. In section 2, we give a brief
overview of the topological 3D gravity with torsion. In section 3, we
derive the supersymmetric extension of this theory with $N=1+1$
gravitini. The derivation is based directly on the form of the
gravitational action and yields a number of relations between coupling
constants, dictated by supersymmetry; the final action agrees with the
result found in \cite{15}. At the end of this section, we give a
concise description of the Chern-Simons formulation of the theory. In
section 4, we start with the standard bosonic AdS asymptotic conditions
and find their natural supersymmetric extension to the gravitini
sector. Then, we obtain the form of the asymptotic parameters and
calculate the related commutator algebra. Section 5 is devoted to the
Hamiltonian analysis and construction of the canonical generator for
local supersymmetry transformations. In section 6, we use the adopted
asymptotic conditions to improve the canonical generators by adding
suitable surface terms. The resulting asymptotic Poisson bracket
algebra is shown to have the form of two super-Virasoro algebras with
different central charges. Section 7 contains concluding remarks, while
appendices summarize some technical details.

Our conventions are essentially the same as in \cite{14}: the Latin
indices  refer to the local orthonormal frame, the Greek indices refer
to the coordinate frame; the middle-alphabet letters
$(i,j,k,\dots,\m,\n,\r,\dots)$ run over $0,1,2$, while
$(\a,\b,\g,\dots)$ run over $1,2$; the metric components in the local
Lorentz frame are $\eta_{ij}=(+,-,-)$; totally antisymmetric tensor
$\ve^{ijk}$ and the related tensor density $\ve^{\m\n\r}$ are both
normalized by $\ve^{012}=+1$; gamma matrices are pure imaginary and
Majorana spinors are real.

\section{3D gravity with torsion} 
\setcounter{equation}{0}

Theory of gravity with torsion can be naturally described as a
Poincar\'e gauge theory (PGT), with an underlying spacetime structure
corresponding to Riemann-Cartan geometry \cite{7,8}.

\prg{PGT in brief.} Basic gravitational variables in PGT are the
triad field $b^i$ and the Lorentz connection $A^{ij}=-A^{ji}$
(1-forms). The corresponding field strengths are the torsion and the
curvature: $T^i=db ^i+A^i{_m}\wedge b^m$,
$R^{ij}=dA^{ij}+A^i{_m}\wedge A^{mj}$ (2-forms). In 3D, we can
simplify the notation by introducing $A^{ij}=:-\ve^{ij}{_k}\om^k$ and
$R^{ij}=:-\ve^{ij}{_k}R^k$, which yields:
\be
T^i=db^i+\ve^i{}_{jk}\om^j\wedge b^k\, ,\qquad
R^i=d\om^i+\frac{1}{2}\,\ve^i{}_{jk}\om^j\wedge\om^k\, .   \lab{2.1}
\ee

Gauge symmetries of the theory are local translations and local
rotations, parametrized by $\xi^\m$ and $\ve^{ij}=:-\ve^{ij}{_k}\th^k$.
In local coordinates $x^\m$, we can write $b^i=b^i{_\m}dx^\m$,
$\om^i=\om^i{_\m}dx^\m$, and local Poincar\'e transformations take the
form
\bea
\d_P b^i{_\m}&=&-\ve^i{}_{jk}b^j{}_{\m}\th^k
   -(\pd_\m\xi^\r)b^i{_\r}-\xi^\r\pd_\r b^i{}_\m \, ,      \nn\\
\d_P\om^i{_\m}&=&-\nabla_\m\th^i-(\pd_\m\xi^\r)\om^i{_\r}
   -\xi^\r\pd_\r\om^i{}_\m \, ,                            \lab{2.2}
\eea
where $\nabla_\m\th^i=\pd_\m\th^i+\ve^i{}_{jk}\om^j{_\m}\th^k$.
The covariant derivative $\nabla=dx^\m\nabla_\m$ acts on a general
tangent-frame spinor/tensor in accordance with its
spinorial/tensorial structure; when $X$ is a form, $\nabla X$ is the
exterior covariant derivative of $X$: $\nabla X=\nabla\wedge X$.

The metric structure of PGT is defined by
$$
g=\eta_{ij}b^i\otimes b^j\equiv g_{\m\n}dx^\m\otimes dx^\n\, ,
\quad \eta_{ij}=\mbox{diag}\,(1,-1,-1)\, .
$$
Metric and connection in PGT are related to each other by the {\it
metricity condition\/}: $\nabla g=0$. Consequently, the geometric
structure of PGT corresponds to {\it Riemann-Cartan geometry\/}.

We display here a useful PGT identity:
\be
\om^i\equiv\tom^i+K^i\, ,                                  \lab{2.3}
\ee
where $\tom^i$ is the Levi-Civita (Riemannian) connection, and $K^i$
is the contortion 1-form, defined implicitly by
$T^i=\ve^i{}_{mn}K^m\wedge b^n$.

\prg{Generalized dynamics.} General gravitational dynamics in
Riemann-Cartan spacetime is determined by Lagrangians which are at most
quadratic in field strengths. Omitting the quadratic terms, we arrive
at the {\it topological\/} Mielke-Baekler (MB) model for 3D gravity
\cite{9}:
\be
I_0=2a\int b^i R_i
    -\frac{\L}{3}\,\int\ve_{ijk}b^i b^j b^k\
    +\a_3I_\cs[\om]+\a_4 \int b^i T_i\, ,                  \lab{2.4}
\ee
where the wedge product sign $\wedge$ is omitted for simplicity. The
first term with $a=1/16\pi G$ is the usual Einstein-Cartan action, the
second term is a cosmological term, $I_\cs[\om]$ is the Chern-Simons
action for the Lorentz connection,
$$
I_\cs[\om]=\int\left(\om^i d\om_i
           +\frac{1}{3}\ve_{ijk}\om^i\om^j\om^k\right)\, ,
$$
and the last term is a torsion counterpart of the first one. The MB
model is a natural generalization of GR with a cosmological constant.

The action $I_0$ defines classical dynamics of the system in the region
{\it outside\/} the gravitational sources through the vacuum field
equations:
\bea
&&2aR_i+2\a_4 T_i-\L\ve_{ijk}b^j b^k=0\, ,                 \nn\\
&&2\a_3 R_i+2aT_i+\a_4\ve_{ijk}b^j b^k=0\, .               \nn
\eea
In the sector $\a_3\a_4-a^2\ne 0$, where the field equations are
non-degenerate, they take the simple form
\be
2T^i=p\ve^i{}_{jk}\,b^j b^k\, ,\qquad
2R^i=q\ve^i{}_{jk}\,b^j b^k\, ,                            \lab{2.5}
\ee
where
$$
p=\frac{\a_3\L+\a_4 a}{\a_3\a_4-a^2}\, ,\qquad
q=-\frac{(\a_4)^2+a\L}{\a_3\a_4-a^2}\, .
$$
Thus, vacuum solutions are characterized by constant torsion and
constant curvature. For $p=0$, the vacuum geometry is Riemannian,
while for $q=0$, it becomes teleparallel.

In Riemann-Cartan spacetime, one can use the identity \eq{2.3} to
express the curvature $R^i=R^i(\om)$ in terms of its {\it
Riemannian\/} piece $\tR^i=R^i(\tom)$ and the contortion. The
resulting identity, combined with the {\it on-shell\/} relation
$K^i=p\,b^i/2$, leads to
\be
\tR^{ij}=-\Leff\,b^i\wedge b^j\, ,\qquad
\Leff:= q-\frac{1}{4}p^2\, ,                               \lab{2.6}
\ee
where $\Leff$ is the effective cosmological constant. The form of
$\tR^{ij}$ implies that our spacetime has maximally symmetric metric:
for $\Leff$ negative, zero or positive, the metric is anti-de Sitter
(AdS), Minkowski or de Sitter, respectively. In what follows, our
attention will be focused on a supersymmetric extension of \eq{2.4}
with {\it negative\/} $\Leff$ (AdS sector).

\section{Supersymmetric extension with \boldmath{$N=1+1$}} 
\setcounter{equation}{0}

In this section, we present a locally supersymmetric extension of 3D
gravity with torsion, based on the action \eq{2.4}. The extension
includes two gravitini fields, compactly denoted as
$\psi^\Pi=(\psi,\psi')$, and is usually referred to as $N=1+1$ AdS
supergravity \cite{2,15}.

\prg{The action.} We start with the action
\be
I=I_0
  -g\int\left(\bar\psi\nabla\psi-im\bar\psi b^i\g_i\psi\right)
  -g'\int\left(\bar\psi'\nabla\psi'
               -im'\bar\psi' b^i\g_i\psi'\right)\, ,       \lab{3.1}
\ee
where $\psi^\Pi=\psi^\Pi_\m dx^\m$ are the gravitini fields (1-forms),
with covariant derivatives
$$
\nabla\psi^\Pi=\Bigl(d-\frac{i}{2}\om^m\g_m\Bigr)\psi^\Pi\, .
$$
Our conventions for the spinors are described in Appendix A. By
construction, the action \eq{3.1} is invariant under the local
Poincar\'e transformations \eq{2.2}, accompanied by
\be
\d_P\psi^\Pi_\m= -\frac{i}{2}\th^k\g_k\psi^\Pi_\m
    -(\pd_\m\xi^\r)\psi^\Pi_\r-\xi^\r\pd_\r\psi^\Pi_\m\, . \lab{3.2}
\ee

\prg{Local supersymmetries.} The action \eq{3.1} is also invariant
under the following local sypersymmetry (SS) transformations with
spinorial parameters $\ve^\Pi=(\ve,\ve')$:
\bea
&&\d_S b^i{}_\m=
  i\bar\ve\g^i\psi_\m +i\bar\ve'\g^i\psi'_\m\, ,          \nn\\
&&\d_S\om^i{}_\m=
  -2im'\bar\ve\g^i\psi_\m-2im\bar\ve'\g^i\psi'_\m\, ,       \nn\\
&&\d_S\psi^\Pi_\m=4a\left(
  \nabla_\m\ve^\Pi-im^\Pi b^k{_\m}\g_k\ve^\Pi\right)\, ,   \lab{3.3}
\eea
where $m^\Pi=(m,m')$. This is true provided the coupling constants
$g,m,g'$ and $m'$ are determined by the following relations (Appendix
B):
\bea
&&2ag=a-2m'\a_3\, ,\qquad 2ag'=a-2m\a_3\, ,                \nn\\
&&2m+\frac{p}{2}=\frac{1}{\ell}\, ,\qquad
  2m'+\frac{p}{2}=-\frac{1}{\ell}\, ,                      \lab{3.4}
\eea
where $\ell$ is the AdS radius, $\ell^{-1}=m-m'$. For $m$ and $m'$
real and different from each other, the effective cosmological
constant is negative:
$$
\Leff=-(m-m')^2\equiv -\frac{1}{\ell^2}<0\, .
$$
The limit $\ell\to\infty$ is not of interest for our present
discussion.

The commutator algebra of the local super-Poincar\'e transformations
\eq{2.2}, \eq{3.2} and \eq{3.3} closes only on shell (Appendix B). In
general, such a situation can be improved by introducing auxiliary
fields. However, since our main goal is to study {\it asymptotic\/}
symmetries, the on-shell terms in the algebra can be safely ignored, as
we shall see in the next section.

\prg{Field equations.} After having constructed the SS action, we can
now derive the field equations. The variation of the action with
respect to $b^i$ and $\om^i$ yields:
\bea
&&2\goth{A}_i:=2aR_i+2\a_4 T_i-\L\ve_{ijk}b^jb^k
   -img\bar\psi\g_i\psi-im'g'\bar\psi'\g_i\psi'=0\, ,      \nn\\
&&2\goth{B}_i:=2aT_i+2\a_3 R_i+\a_4 \ve_{ijk}b^jb^k
   -\frac{1}{2}ig\bar\psi\g_i\psi
   -\frac{1}{2}ig'\bar\psi'\g_i\psi'=0\, .                 \lab{3.5}
\eea For $\a_3\a_4-a^2\ne 0$, and taking into account \eq{3.4},
these equations can be rewritten as \bsubeq\lab{3.6} \bea
&&2F_i:=2T_i-\frac{i}{4a}\bar\psi\g_i\psi
  -\frac{i}{4a}\bar\psi'\g_i\psi'-p\ve_{ijk} b^j b^k=0\, , \nn \\
&&2G_i:=2R_i+\frac{im'}{2a}\bar\psi\g_i\psi
  +\frac{im}{2a}\bar\psi'\g_i\psi'
  -q\ve_{ijk} b^j b^k=0\, .                                \lab{3.6a}
\eea
The variations with respect to $\bar\psi$ and $\bar\psi'$ yields the
gravitini field equations:
\bea
&&H:=\nabla\psi-imb^i\g_i\psi=0\, ,                        \nn\\
&&H':=\nabla\psi'-im'b^i\g_i\psi'=0\, .                    \lab{3.6b}
\eea
\esubeq

Before continuing, we introduce the compact notation for the
multiplets of fields and field equations:
\be
\Phi^A=(b^i,\om^i,\psi,\psi')\, ,\qquad
E^A=(F^i,G^i,H,H')\, .                                     \lab{3.7}
\ee

\prg{Chern-Simons formulation.} It is well known that the original MB
model can be described as a Chern-Simons (CS) gauge theory
\cite{11,13}. The extension of this important property of 3D gravity
with torsion to the supersymmetric action \eq{3.1} is almost evident.
To show that, we begin by recalling the identity \cite{11}
\be
\cL_0(b,\om)+ad(b^i\om_i)
            \equiv\k_-\cL_\cs(A^-)-\k_+\cL_\cs(A^+)\, ,    \lab{3.8}
\ee
which relates the bosonic MB model \eq{2.4} with two independent
$SL(2,R)$ CS theories. Here,
\bea
&&\cL_\cs(A)=A^i dA_i+\frac{1}{3}\ve_{ijk}A^i A^j A^k\, ,  \nn\\
&&A^{-i}=\om^i+2mb^i\, ,\qquad  A^{+i}=\om^i{}+2m'b^i\, ,  \nn\\
&&\k_-=\ell ag\, ,\qquad \k_+=\ell ag'\, .                 \nn
\eea
The new dynamical variables $A^{\mp i}$ (1-forms) are $SL(2,R)$ gauge
potentials.

It is of particular importance to note that the gauge content of 3D
gravity with torsion can be interpreted in two equivalent ways:
\bitem
\item[(a)] as a PGT \cite{8} (section II), or \vsm
\item[(b)] as an ordinary  gauge theory, based on the AdS group
$SO(2,2)$ for $\Leff<0$ \cite{3}.
\eitem
Since $SO(2,2)$ is locally isomorphic to $SL(2,R)\times SL(2,R)$, the
latter interpretation makes the relation to the CS structure more
transparent.

Using the identity \eq{3.8}, one easily finds that the supergravity
action \eq{3.1} can be rewritten in the form
\be
I+a\int d(b^i\om_i)
  =\k_-I_\cs[A^-,\psi]-\k_+I_\cs[A^+,\psi']\, ,            \lab{3.9}
\ee
where each of the two terms on the rhs is an $Osp(1|2)$ CS action
\cite{19,16},
\be
I_\cs[A^\mp,\psi]=I_\cs[A^\mp] \mp\frac{1}{\ell a}\int\bar\psi D\psi\, ,
\ee
with $D\psi:=d\psi-\frac{1}{2}iA^k\g_k\psi$. Thus, the complete
$N=1+1$ supergravity theory \eq{3.1} can be viewed as a super-CS
gauge theory, based on $Osp(1|2)\times Osp(1|2)$ \cite{15}. This
explains the notation $N=1+1$ for the number of gravitini and the
name AdS supergravity.

Following the interpretation (b), one can find the form of
$Osp(1|2)\times Osp(1|2)$ gauge transformations $\d\Phi^A$ on the
supergravity multiplet $\Phi^A$. The related commutator algebra closes
{\it off shell\/}. On the other hand, in agreement with (a), one can
define a new set of local parameters, such that $\d\Phi^A$ takes the
super-Poincar\'e form {\it on shell\/}:
\be
\d\Phi^A_\m=\d_{SP}\Phi^A_\m -\xi^\r E^A_{\m\r}\, .        \lab{3.11}
\ee
The canonical version of these results is derived in the next section.

The CS structure of 3D gravity greatly simplifies its dynamical
content \cite{3}. In the present approach, however, our attention is
focused directly on the supergravity action \eq{3.1}.

\section{Asymptotic conditions} 
\setcounter{equation}{0}

General principles underlying the choice of asymptotic conditions for
the supergravity theory \eq{3.1} are the same ones as those formulated
two decades ago in \cite{1}: we assume that the asymptotic set of
fields is general enough to include the black hole configuration (with
$\psi=\psi'=0$), together with all its transforms under the action of
global $Osp(1|2)\times Osp(1|2)$, but sufficiently regular to yield
well-defined canonical generators. These principles lead to a choice
that ensures the existence of a nontrivial asymptotic symmetry---the
super-conformal symmetry.

\subsection{Asymptotic configuration of fields}

For $\Leff<0$, the asymptotic structure of 3D gravity with torsion is
well understood \cite{13}. In order to generalize the corresponding
asymptotic symmetry in a natural way---from conformal to
super-conformal---the bosonic fields in the supergravity theory
\eq{3.1} should keep their asymptotics unchanged. In
Schwartzschield-like coordinates $x^\m=(t,r,\vphi)$, this means that
\bsubeq\lab{4.1}
\be
b^i{_\m}= \left( \ba{ccc}
   \dfrac{r}{\ell} +\cO_1 & \cO_4  & \cO_1  \\
   \cO_2 & \dfrac{\ell}{r}+\cO_3   & \cO_2  \\
   \cO_1 & \cO_4                   & r+\cO_1
                 \ea
          \right)   \, ,                                   \lab{4.1a}
\ee
where $\cO_n$ are terms that tend to zero as $r^{-n}$ or faster, and
\be
\om^i{_\m}=\left( \ba{ccc}
   \dfrac{pr}{2\ell}+\cO_1 & \cO_2  & -\dfrac{r}{\ell}+\cO_1 \\
   \cO_2 & \dfrac{p\ell}{2r}+\cO_3  & \cO_2                  \\
   -\dfrac{r}{\ell^2}+\cO_1 & \cO_2 & \dfrac{pr}{2}+\cO_1
                  \ea
                  \right)  \, .                            \lab{4.1b}
\ee
Note that in any $\cO_n=c/r^n$, we assume that $c=c(t,\vphi)$.

In the next step, we have to make an appropriate choice for the
fermionic fields. This can be done following the ideas expressed by
the above principles, as in \cite{1,16}, but we find it practically
simpler to rely on another general principle:
\bitem
\item[\gp] The expressions than vanish on shell should have an
arbitrarily fast asymptotic decrease, as no solutions of the field
equations are thereby lost.
\eitem
The application of this principle to the field equations \eq{3.6a}
yields $F^i=\hcO$ and $G^i=\hcO$, where $\hcO$ is a term with
arbitrarily fast asymptotic decrease. Combining these relations with
the asymptotic conditions \eq{4.1} leads to the preliminary results
for $\psi$ and $\psi'$:
\bea
&&\psi_0,\psi_2=\hcO_{1/2}\, ,\qquad \psi_1=\hcO_{5/2}\, , \nn\\
&&\psi'_0,\psi'_2=\hcO_{1/2}\, ,\qquad\psi'_1=\hcO_{5/2}\,.\nn
\eea
As described in Appendix C, this result in combination with
\eq{4.1a}, \eq{4.1b} and $H=\hcO$, leads to the improved asymptotic
form of $\psi$:
\bea
&&\ell\psi_0=\left(\ba{c}
             \ell r^{-1/2}\chi^-(x^-)+\cO_{5/2} \\
                   \cO_{3/2}
                   \ea\right)=-\psi_2\, ,                  \nn\\
&&\psi_1=\left(\ba{c}
               \cO_{7/2} \\
                \ell^2 r^{-5/2}\ups^-(t,\vphi)+\cO_{9/2}
               \ea\right)\, ,                              \lab{4.1c}
\eea
where $\chi^-,\ups^-$ are Grassmann odd functions. Similar arguments
lead to:
\bea
&&\ell\psi'_0=\left(\ba{c}
                \cO_{3/2} \\
                \ell r^{-1/2}\chi^+(x^+)+\cO_{5/2}
                \ea\right)=\psi'_2\, ,                     \nn\\
&&\psi'_1=\left(\ba{c}
                \ell^2 r^{-5/2}\ups^+(t,\vphi)+\cO_{9/2} \\
                \cO_{7/2}
                \ea\right)\, .                             \lab{4.1d}
\eea
\esubeq
The light-cone components of $X^\m$ are defined as $X^\pm=X^0/\ell\pm
X^2$.

We adopt \eq{4.1} as the asymptotic conditions for the supergravity
theory \eq{4.1}. Technically, it is convenient to rewrite these
conditions in the form
\bea
&& b^i{_\m}=\langle b^i{_\m}\rangle + B^i{_\m}\, ,\qquad
  \om^i{_\m}=\langle\om^i{_\m}\rangle +\Om^i{_\m}\, ,      \nn\\
&& \psi_\m=\langle\psi_\m\rangle + \Psi_\m\, ,\qquad
   \psi'_\m=\langle\psi'_\m\rangle +\Psi'_\m\, ,           \nn
\eea
where $\langle\Phi^A\rangle$ are leading-order, and
$(B^i{_\m},\Om^i{_\m},\Psi_\m,\Psi'_\m)$ higher-order terms.

\subsection{Asymptotic parameters}

Now, we wish to find the most general subset of local super-Poincar\'e
transformations that preserves the asymptotic conditions \eq{4.1}.
Technical derivation is described in Appendix D.

As a result, we find that the asymptotic form of the spinorial
parameters is:
\bsubeq\lab{4.2}
\be
\ve=\left(\ba{c}
          \ell r^{-1/2}\pd_-\eps^-+\cO_{5/2} \\
          r^{1/2}\eps^-+\cO_{3/2}
          \ea\right)\, ,\qquad
\ve'=\left(\ba{c}
           r^{1/2}\eps^++\cO_{3/2} \\
           \ell r^{-1/2}\pd_+\eps^++\cO_{5/2}
           \ea\right)\, .
\ee
where $\eps^\mp=\eps^\mp(x^\mp)$ are Grassmann odd parameters of the
residual symmetry. Local Poincar\'e parameters $\xi^\m$ are found to
have the following asymptotic form:
\bea
&&\frac{1}{\ell}\xi^0=T+\frac{\ell^4}{2r^2}\frac{\pd^2 T}{\pd t^2}
  -\frac{i\ell^2}{2r^2}\left[\eps^-(\chi^-+\ups^-)
  +\eps^+(\chi^++\ups^+)\right]+\cO_4\, ,                  \nn\\
&&\xi^2=S-\frac{\ell^2}{2r^2}\frac{\pd^2 S}{\pd\vphi^2}
  -\frac{i\ell^2}{2r^2}\left[\eps^-(\chi^--\ups^-)
  -\eps^+(\chi^+-\ups^+)\right]+\cO_4\, ,                  \nn\\
&&\xi^1=-\ell r\frac{\pd T}{\pd t}+\cO_1\, .
\eea
Note that the soul of $\xi^\m$ \cite{20}, which is an even
combination of Grassman odd quantities, is present only in the
subleading terms. Similarly:
\bea
&&\th^0=-\frac{\ell}{r}\frac{\pd^2 S}{\pd\vphi^2}
        -\frac{i\ell}{r}\left(\eps^-\chi^-
        -\eps^+\chi^+\right)+\cO_3\, ,                     \nn\\
&&\th^2=\frac{\ell^3}{r}\frac{\pd^2 T}{\pd t^2}
        -\frac{i\ell}{r}\left(\eps^-\chi^-
        +\eps^+\chi^+\right)+\cO_3\, ,                     \nn\\
&&\th^1=\frac{\pd T}{\pd\vphi}+\cO_2\, .
\eea
\esubeq
Here, $T$ and $S$ are parameters that satisfy the conditions $\pd_\pm
T^\mp=0$, where $T^\mp:=T\mp S$.

Equations \eq{4.2} imply that the full residual symmetry is
characterized by four chiral parameters: $T^\mp(x^\mp)$ and
$\eps^\mp(x^\mp)$.

\subsection{The asymptotic commutator algebra}

Local structure of the asymptotic symmetry is determined by the form
of its commutator algebra. The super-Poincar\'e gauge algebra is
closed, but only {\it on shell\/} (Appendix B):
\be
[\d_{SP}(1),\d_{SP}(2)]=\d_{SP}(3)
                        +\cO(\mbox{field equations})\, .   \lab{4.3}
\ee
where $\d_{SP}(1)=\d_{SP}(\xi^\m_1,\th^i_1,\ve_1,\ve'_1)$, etc. In
such a case, one usually needs to introduce auxiliary fields.
However, the general principle \gp\ ensures that the term
$\cO(\mbox{field equations})$ in \eq{4.3} can be safely ignored in
the {\it asymptotic\/} region. Hence, we can rely on the
super-Poincar\'e algebra as a respectable tool for studying
asymptotic symmetries.

If we substitute the restricted gauge parameters \eq{4.2} into the
composition law \eq{B4b}, the leading-order terms define the
composition law for the asymptotic parameters:
\bea
&&T^\mp_3=T_1^\mp\pd_\mp T_2^\mp-T_2^\mp\pd_\mp T_1^\mp
  + 8ia\eps^\mp_1\eps^\mp_2\, ,                          \nn\\
&&\eps^\mp_3=T_1^\mp\pd_\mp\eps^\mp_2
  -\frac{1}{2}(\pd_\mp T_1^\mp )\eps^\mp_2-T_2^\mp\pd_\mp\eps^\mp_1
  +\frac{1}{2}(\pd_\mp T_2^\mp)\eps^\mp_1\, .              \lab{4.4}
\eea
Two restricted gauge transformations with the same $(T^\mp,\eps^\mp)$
differ by a pure gauge transformation. In order to have a consistent
commutator algebra in the set of restricted gauge transformations, we
have to factor our pure gauge transformations. The resulting quotient
algebra defines the algebraic structure of the asymptotic symmetry
\cite{21,22,16}.

The composition law \eq{4.4} defines an asymptotic symmetry algebra,
which can be expressed in a more familiar form by using the Fourier
expansion of $T^\mp,\eps^\mp$. Indeed, with
$$
\d_{SP}(T^\mp=a_n^\mp e^{inx^\mp})=:a_n\ell^\mp_n\, ,\qquad
\d_{SP}(\eps^\mp=\a_n^\mp e^{inx^\mp})=:i\a^\mp_n q^\mp_n\sqrt{4a}\,,\nn
$$
the asymptotic algebra is expressed in terms of two independent
super-Virasoro algebras with vanishing central charges:
\bea
i\left[\ell_n^\mp,\ell_m^\mp\right]&=&(n-m)\ell_{m+n}^\mp\,,\nn\\
i\left[\ell_n^\mp,q_m^\mp\right]
   &=&\left(\frac{1}{2}n-m\right)q^\mp_{m+n}\, ,           \nn\\
i\left\{q_n^\mp,q^\mp_m\right\}&=& 2\ell^\mp_{n+m}\, .  \lab{4.5}
\eea

In the sectors with periodic or anti-periodic boundary conditions for
fermions \cite{4}, index of $q^\mp_m$ takes on integer or
half-integer values, respectively.

\section{Hamiltonian analysis} 
\setcounter{equation}{0}

Gauge symmetries of a dynamical system are best understood in the
canonical formalism. In particular, Hamiltonian analysis is a powerful
tool for exploring asymptotic conditions and their symmetries
\cite{8,14}.

\subsection{Hamiltonian and constraints}

Starting with the basic Lagrangian variables
$(b^i{}_\m,\om^i{}_\m,\psi_\m,\psi'_\m )$, we denote their canonical
momenta by $(\pi^i{}_\m,\Pi^i{}_\m,\bar\pi^\m,\bar\pi'^\m)$. Since
$\psi_\m$ and $\psi'_\m$ are fermionic (Grassmann odd) fields, we
define the canonical momenta by left derivatives of the Lagrangian and
use the generalized Poisson brackets (PB) \cite{23}. The primary
constraints are:
\be
\ba{l}
\phi_i{}^0:=\pi_i{}^0\approx 0\, , \\[3pt]
\phi_i{}^\a:=\pi_i{}^\a-\a_4\ve^{0\a\b}b_{i\b}\approx 0\, , \\[3pt]
\bar f^0:=\bar\pi^0\approx 0\, , \\[3pt]
\bar f'^{0}:=\bar\pi'^{0}\approx 0\, ,
\ea \,
    \ba{l}
    \Phi_i{}^0:=\Pi_i{}^0\approx 0\, , \\[3pt]
    \Phi_i{}^\a:=\Pi_i{}^\a-\ve^{0\a\b}\left(2a b_{i\b}
                 +\a_3\om_{i\b}\right)\approx 0\, , \\[3pt]
    \bar f^\a :=\bar\pi^{\a}
                -g\ve^{0\a\b}\bar\psi_{\b}\approx 0\, , \\[3pt]
    \bar f'^{\a}:=\bar\pi'^{\a}
                  -g'\ve^{0\a\b}\bar{\psi'}_{\b}\approx 0\, .
    \ea                                                    \lab{5.1}
\ee
The canonical Hamiltonian is linear in unphysical variables, as
expected:
$$
\cH_c= b^i{}_0\cH_i+\om^i{}_0\cK_i+\bar\psi_0\cF
      +\bar\psi'_0\cF'+\pd_\a D^\a\, ,
$$
where
\bea
&&\cH_i=-\ve^{o\a\b}\goth{A}_{i\a\b}\, ,\qquad
  \cK_i=-\ve^{o\a\b}\goth{B}_{i\a\b}\, ,                   \nn\\
&&\cF=2g\ve^{0\a\b}H_{\a\b}\, ,\qquad
  \cF'=2g'\ve^{0\a\b}H'_{\a\b}\, ,                         \nn\\
&&D^\a=\ve^{0\a\b}\Bigl[ \om^i{}_0\left( 2ab_{i\b}
   +\a_3\om_{i\b}\right)+\a_4b^i{}_0 b_{i\b}
   -g\bar\psi_\b\psi_0-g'\bar\psi'_\b\psi'_0\Bigr]\, ,     \nn
\eea
and $(\goth{A}_{i\a\b},\goth{B}_{i\a\b})$ and
$(H_{\a\b},H'_{\a\b})$ are spatial components of the field
equations \eq{3.5} and \eq{3.6b}, respectively. Going over to the
total Hamiltonian,
\be
\cH_T=\cH_c+u^i{}_\m\phi_i{}^\m+v^i{}_\m\Phi_i{}^\m
   -\bar f^\m w_\m-\bar f'^{\m }w'_{\m }\, ,               \lab{5.2}
\ee
we find that the consistency conditions of the sure primary
constraints $\pi_i{}^0$, $\Pi_i{}^0$, $\bar\pi^{0}$ and
$\bar\pi'^{0}$ yield the secondary constraints:
\be
\cH_i\approx0\, , \qquad \cK_i\approx0\, , \qquad
\cF\approx 0\, ,\qquad \cF'\approx 0\, .                   \lab{5.3}
\ee

The consistency of the additional primary constraints $\phi^i{}_\a$,
$\Phi^i{}_\a$, $\bar f^\a$ and  $\bar f'^\a$ leads to the
determination of the multipliers $u^i{}_\b$, $v^i{}_\b$, $w_\a $ and
$w'_\a$:
$$
\underbar{F}{\,}_{i0\b}\approx 0\, , \qquad
\underbar{G}{\,}_{i0\b}\approx 0\, ,\qquad
\underbar{H}{\,}_{0\b}\approx 0\, ,\qquad
\underbar{H}'{}_{0\b}\approx 0\, ,
$$
where the underbars in $\underbar{F}$, $\underbar{G}$, $\underbar{H}$
and $\underbar{H}'$ mean that the terms $\dot b^i{}_\b$,
$\dot\om^i{}_\b$, $\dot\psi_\a$ and $\dot\psi'_\a$ are replaced with
$u^i{}_\b$, $v^i{}_\b$, $ w_\b$ and $w'_\b$. On shell, these
relations reduce to the $0\b$ components of the field equations
\eq{3.6}.

The substitution of the determined multipliers into \eq{5.2}
yields the final form of the total Hamiltonian:
\bsubeq
\bea
\cH_T&=&\hcH_T+\pd_\a\hat{D}^\a\, ,                        \nn\\
\hcH_T&=&b^i{}_0\hcH_i+\om^i{}_0\hat\cK_i
  +\bar\psi_0\hcF+\bar\psi'_0\hcF'                         \nn\\
&&+u^i{}_0\pi_i{}^0+v^i{}_0\Pi_i{}^0
         -\bar\pi^0w_0-\bar\pi'^0w'_0\, ,                  \lab{5.4a}
\eea
where
\bea
&&\hcH_i=\cH_i-\nabla_\b\phi_i{}^\b
  +\ve_{ijk}b^j{}_\b\left(p\phi^{k\b}+q\Phi^{k\b}\right)
  -im\bar f^\a\g_i\psi_\a-im'\bar f'^\a \g_i\psi'_\a \, ,  \nn\\
&&\hcK_i=\cK_i-\nabla_\b\Phi_i{}^\b
  -\ve_{ijk}b^j{}_\b\phi^{k\b}-\frac{i}{2}\bar f^\a \g_i\psi_\a
  -\frac{i}{2}\bar f'^\a\g_i\psi'_\a \, ,                  \nn\\
&&\hcF=\cF+\frac{i}{4a}\g^n\psi_\a \phi_n{}^\a
  -\frac{im'}{2a}\g^n\psi_{\a}\Phi_n{}^\a
  +\nabla_\a f^{\a}+imb^n{}_\a\g_nf^\a\, ,                 \nn\\
&&\hcF'=\cF'+\frac{i}{4a}\g^n\psi'_\a\phi_n{}^\a
  -\frac{im}{2a}\g^n\psi'_\a\Phi_n{}^\a
  +\nabla_\a f'^\a +im'b^n{}_\a\g_nf'^\a\, ,               \nn\\
&&\hat D^\a=b^n{}_0\pi_n{^\a}+\om^n{}_0\Pi_n{}^\a
  -\bar\pi^\a\psi_0-\bar\pi'^\a\psi'_0\, .                 \lab{5.4b}
\eea
\esubeq

Further investigation of the consistency procedure is facilitated by
observing that the secondary constraints $\hcH_i,\hcK_i,\hcF,\hcF'$
obey the PB relations \eq{E3}. Consequently, the consistency
conditions of the secondary constraints  are identically satisfied,
and the Hamiltonian consistency procedure is thereby completed.

Final classification of the constraints is summarized in the
following table.
\begin{center}
\doublerulesep 1.8pt
\begin{tabular}{lll}
\multicolumn{3}{l}{\hspace{16pt}Table 1. Classification
                                         of constraints} \\
                                                      \hline\hline
\rule{0pt}{12pt}
&~First class\phantom{x}&~Second class\phantom{x}     \\
                                                      \hline
\rule[-1pt]{0pt}{15pt}
\phantom{x}Primary &~$\p_i{^0},\Pi_i{^0},\bar\pi^0,\bar\pi'^0$
            &~$\phi_i{^\a},\Phi_i{}^{\a},\bar f^\a,\bar f'^\a$ \\
                                                      \hline
\rule[-1pt]{0pt}{15pt}
\phantom{x}Secondary\phantom{x} &~$\hcH_i,\hcK_i,\hcF,\hcF'$
                                &~                    \\
                                                      \hline\hline
\end{tabular}
\end{center}

\prg{Asymptotics of the phase space.} In order to extend the
asymptotic conditions \eq{4.1} to the whole phase space, one should
complete the procedure by choosing an appropriate asymptotic behavior
for the the momentum variables, too. This is easily done by applying
the principle \gp\ from the previous section to the primary
constraints \eq{5.1}. As we shall see, the resulting phase-space
structure ensures the existence of well-defined canonical generators.

\subsection{The canonical generator}

Having completed the Hamiltonian analysis, we can now construct the
canonical generator of gauge transformations \cite{24}. It has the form
\bsubeq\lab{5.5}
\be
G=\hG_1+\hG_2+\hG_3+\hG_4\, ,
\ee
where
\bea
&&\hG_1:=\dot{\t^i}\pi_i{}^0
  +\t^i\left[\hcH_i -\ve_{ijk}\left(\om^j{}_0
  -pb^j{}_0\right)\pi^{k0}+q\ve_{ijk}b^j{}_0\Pi^{k0}
  -im\bar\pi^0\g_i\psi_0-im'\bar\pi'^0\g_i\psi'_0\right]\,,\nn\\
&&\hG_2:=\dot{\s^i}\Pi_i{}^0+\s^i\left[\hcK_i
  -\ve_{ijk}\left(b^j{}_0\pi^{k0}+\om^j{}_0\Pi^{k0}\right)
  -\frac{i}{2}\bar\pi^0\g_i\psi_0
  -\frac{i}{2}\bar\pi'^0\g_i\psi'_0\right]\, ,             \nn\\
&&\hG_3:=-\bar\pi^0\dot{\a}+\bar\a\left[\hcF
  -\frac{i}{2}\left(\om^n{}_0+2mb^n{}_0\right)\g_n\pi^0
  +\frac{i}{4a}\left(\pi_n{}^0
                  -2m'\Pi_n{}^0\right)\g^n\psi_0\right]\, ,\nn\\
&&\hG_4:=-\bar\pi'^0\dot{\a}'+\bar\a'\left[\hcF'
  -\frac{i}{2}\left(\om^n{}_0+2m'b^n{}_0\right)\g_n\pi'^0
  +\frac{i}{4a}\left(\pi_n{}^0
                  -2m\Pi_n{}^0\right)\g^n\psi'_0\right] \, .
\eea
The action of $\hG$ on the fields is defined  by the PB operation
$\bar\d\Phi^A=\{\Phi^A,G\}$:
\bea
&&\bar\d b^i=\ve^i_{jk}b^j\s^k+\nabla\t^i
  -p\ve^i{}_{jk}b^j\t^k+\frac{i}{4a}\bar\a\g^i\psi
  +\frac{i}{4a}\bar\a'\g^i\psi'\, ,                        \nn\\
&&\bar\d\om^i=\nabla\s^i-q\ve^i{}_{jk}b^j\t^k\
  -\frac{im'}{2a}\bar\a\g^i\psi-\frac{im}{2a}\bar\a'\g^i\psi'\,,\nn\\
&&\bar\d\psi=\frac{i}{2}\left(\s^n+2m\t^n\right)\g_n\psi
  +\nabla\a-imb^n\g_n\a\, ,                                \lab{5.5c}
\eea
\esubeq
and $\bar\d\psi'$ is the ``primed" version of $\bar\d\psi$, with
$(m,\a,\psi)\to(m',\a',\psi')$. The following properties of the
canonical transformations \eq{5.5c} clarify their meaning:
\bitem
\item[--] the $\s^i$-transformations are local Lorentz
transformations, \vsm
\item[--] the $(\a,\a')$-transformations are local SS
transformations, \vsm
\item[--] the set of transformations generated by $\hG_1+(p/2)\hG_2$
with $\s^i=\t^i$, coincides with the ``extra local symmetry" introduced
in Eq. (3.5) of Ref. \cite{15}.
\eitem
However, the most complete information is given by the following
statement:
\bitem
\item[(a)] the canonical transformations \eq{5.5c} are ordinary
$osp(1|2)\times osp(1|2)$ gauge transformations, defined by the
parameters
$$
u^{-i}=-\s^i-2m\t^i\, ,\quad u^{+i}=-\s^i-2m'\t^i\, ,\quad
(\ve,\ve')=\frac{1}{4a}(\a,\a')\, .
$$
\eitem

Local translations are hidden in \eq{5.5c} \cite{2,14}. To disclose
them, we introduce the set of new parameters
\bea
&&\t^i=-\xi^\r b^i{}_\r\, ,\qquad
  \s^i=-(\th^i+\xi^\r\om^i{}_\r)\, ,                       \nn\\
&&\a=4a\ve-\xi^\r\psi_{\r }\, ,\qquad
  \a'=4a\ve'-\xi^\r\psi'_{\r }\, ,                         \lab{5.6}
\eea
whereupon \eq{5.5c} takes the form of local super-Poincar\'e
transformations {\it on shell}, in accordance with \eq{3.11}.
Expressed in terms of the new parameters, the generator $G$ reads:
\bea
&&G=-G_1-G_2-G_3-G_4 \, ,                                  \lab{5.7}\\
&&G_1:=\dot\xi^\r\left(b^i{}_\r\pi_i{}^0
  +\om^i{}_\r\Pi_i{}^0-\bar\pi^0\psi_\r
  -\bar\pi'^0\psi'_\r\right)                               \nn\\
&&\qquad\,
  +\xi^\r\left[b^i{}_\r\hat\cH_i+\om^i{}_\r\hat\cK_i
  +\bar\psi_\r\hat\cF+\bar\psi'_\r \hat\cF'+(\pd_\r b^i_0)\pi_i{}^0
  +(\pd_\r\om^i{}_0)\Pi^i{}_0 -\bar\pi^0\pd_\r\psi_0
  -\bar\pi'^0\pd_\r\psi'_0\right],                         \nn\\
&&G_2:=\hG_2(\s^i\to\th^i)\, ,                             \nn\\
&&G_3:=-4a\hG_3(\a\to\ve)\, ,\qquad
  G_4:=-4a\hG_4(\a'\to\ve')\, .                            \nn
\eea
Here, the time derivatives $\dot b^i{}_\m$, $\dot\om^i{}_\m$,
$\dot{\psi}_\m$ and $\dot{\psi'}_\m$ are shorts for $u^i{}_\m$,
$v^i{}_\m$, $w_\m$ and $w'_\m$, respectively. As we discussed in the
previous section, the super-Poincar\'e generator \eq{5.7} is quite
suitable for the analysis of asymptotic symmetries.

In the above expressions, the integration symbol $\int d^2 x$ is
omitted for simplicity; later, when necessary, it will be restored.

\section{Canonical realization of the asymptotic symmetry} 
\setcounter{equation}{0}

In this section, we study the influence of the adopted asymptotic
conditions on the canonical structure of the theory: we construct the
improved gauge generators, examine their asymptotic canonical algebra
and prove the conservation laws.

\subsection{Improving the canonical generator}

The canonical generator acts on dynamical variables via the PB
operation, which is defined in terms of functional derivatives. A
phase-space functional $F=\int d^2x f(\phi,\pd\phi,\pi,\pd\pi)$ has
well-defined functional derivatives if its variation can be written
in the form $\d F=\int d^2x\left[A(x)\d\phi(x)+B(x)\d\pi(x)\right]$.
In order to ensure this property for our generator \eq{5.7}, we have
to improve its form by adding certain surface terms \cite{21}.

Under the adopted asymptotic conditions, the fermionic pieces of
$G_1$ and $G_2$ are regular, hence the form of the corresponding
surface terms remains the same as in the bosonic sector of the theory
\cite{14}: $G_2$ is regular, and
\bsubeq
\bea
&&\tG_1=G_1-\G_B\, ,                                       \nn\\
&&\G_B:=-\int_0^{2\pi}d\vphi\left(\ell T\cE^1+S\cM^1\right)\, ,
\eea
where
\bea
&&\cE^\a:=
  2\ve^{0\a\b}\left[\left(a+\frac{\a_3p}{2}\right)\om^0{}_\b
  +\left(\a_4+\frac{ap}{2}\right)b^0{}_\b+\frac{a}{\ell}b^2{}_\b
  +\frac{\a_3}{\ell}\om^2{}_\b\right]b^0{}_0\, ,           \nn\\
&&\cM^\a:=
  -2\ve^{0\a\b}\left[\left(a+\frac{\a_3p}{2}\right)\om^2{}_\b
  +\left(\a_4+\frac{ap}{2}\right)b^2{}_\b+\frac{a}{\ell}b^0{}_\b
  +\frac{\a_3}{\ell}\om^0{}_\b\right]b^2{}_2\, .           \nn
\eea
For the generators $G_3$ and $G_4$, we obtain:
\bea
&&\tG_3+\tG_4=G_3+G_4-\G_F\, ,                             \nn\\
&&\G_F:=-8a\int_0^{2\pi} d\vphi
        \left(g\bar\ve\psi_2+g'\bar\ve'\psi'_2\right)
  =-8i\ell a\int_0^{2\pi}d\vphi(g\eps^-\chi^--g'\eps^+\chi^+)\, .
\eea
\esubeq
Thus, the complete improved  gauge generator is given by
\be
\tG=G+\G\, ,\qquad \G:=\G_B+\G_F\, .
\ee
The adopted asymptotic conditions guarantee that $\tG$ is finite and
differentiable functional. The boundary term depends only on the
asymptotic parameters $T^\mp$ and $\eps^\mp$.

\subsection{Canonical algebra}

In the canonical formalism, the asymptotic symmetry is described by
the PB algebra of the improved generators $\tG$. Using the notation
$\tG_{(1)}:=\tG[T^-_1,T^+_1,\eps^-_1,\eps^+_1]$, and so on, and
relying on the general method described in \cite{14}, one finds that
the PB algebra has the form
\bsubeq\lab{6.3}
\be
\left\{\tG_{(2)},\tG_{(1)}\right\}=\tG_{(3)}+C_{(3)}\, ,   \lab{6.3a}
\ee
where $T_3^\mp,\eps_3^\mp$ are defined by the composition law
\eq{4.4}, and $C_{(3)}$ is the classical {\it central term\/}:
\bea
C_{(3)}&=&2\ell ag\int_0^{2\pi}d\vphi\left[
   (\pd_-^2 T_1^-)(\pd_- T_2^-)-16ia\eps^-_2\pd_-^2\eps^-_1\right]\nn\\
&&+2\ell ag'\int_0^{2\pi}d\vphi\left[
   (\pd_+^2 T_1^+)(\pd_+T_2^+)-16ia\eps^+_2\pd_+^2\eps^+_1\right]\, .
\eea
\esubeq
The bosonic piece of $C_{(3)}$ is the same as calculated in
\cite{14}.

\prg{Conserved charges.} For $\xi^0=1$, the improved generator $\tG$
reduces to $-\tilde H_T$, so that the PB algebra
\eq{6.3} implies its conservation:
\bea
\frac{d}{dt}\tG&=&\ell\frac{\pd}{\pd t}\tG
  +\left\{\tG,\,\tilde{H}_T\right\}                        \nn\\
&\approx&\frac{\pd}{\pd t}\G[T,S,\eps^-,\eps^+]
  -\frac{1}{\ell}\G[\pd_2 S,\pd_2 T,\pd_-\eps^-,\pd_+\eps^+]=0\,.\nn
\eea
This result also implies the conservation of the boundary term $\G$.
For any specific solution of the theory, the values of these boundary
terms define conserved charges for that solution. In particular, energy
and angular momentum are defined as the values of $-\G$ for $\ell T=1$
(time translations) and $S=1$ (rotations), respectively \cite{14}, and
supercharges are defined by the values of $-\G$ for constant asymptotic
parameters $\eps^\mp$: $i\eps^\mp Q^\mp=-\G$.

For the black hole solution with zero gravitini fields, energy and
angular momentum have the same values as in the bosonic sector
\cite{14}, while the supercharges vanish.

\prg{Central charge.} After introducing the Fourier modes
\bea
&&a_n^\mp L_n^\mp:=-\tG(T^\mp=a_n^\mp e^{inx^\mp})\, ,     \nn\\
&&i\a_n^\mp Q_n^\mp:=-\tG(\eps^\mp=\a_n^\mp e^{inx^\mp})/\sqrt{4a}\,,\nn
\eea
the PB algebra \eq{6.3} takes a more familiar form: it is expressed
in terms of two independent super-Virasoro algebras with different
central charges:
\bsubeq\lab{6.4}
\bea
i\{L^\mp_n,L^\mp_m\}&=&(n-m)L^\mp_{n+m}
                      +\frac{c^\mp}{12}n^3\d_{m+n,0}\, ,   \nn\\
i\{L^\mp_n,Q^\mp_m\}&=&
   \left(\frac{1}{2}n-m\right)Q^\mp_{m+n}\, ,              \nn\\
 i\{Q_n^\mp,Q_m^\mp\}&=&
   2L^\mp_{n+m}+\frac{c^\mp}{3}n^2\d_{m+n,0}\, ,           \lab{6.4a}
\eea
where
\be
c^-=12\cdot 4\pi\ell ag\, ,\qquad c^+=12\cdot 4\pi\ell ag'\, .
\ee
\esubeq
The canonical algebra \eq{6.4} is the central extension of the
commutator algebra \eq{4.5}.

In a spacetime with the black hole topology, the fermions can be
periodic (index of $Q_m$ integer) or antiperiodic (index of $Q_m$
half-integer) \cite{4}. In the first case, there is a subalgebra
generated by $(L_0,Q_0)^\mp$, for which the central charge vanishes;
in the second, after shifting $L_0^\mp$ by $c^\mp/24$, one finds that
the central charge vanishes on the subalgebra $osp(1|2)\times
osp(1|2)$, generated by $(L_{\mp 1},L_0,Q_{\mp 1/2})^\mp$. In the
context of Riemannian geometry with $c^-=c^+=3\ell/2G$, these global
symmetries are interpreted as exact symmetries of the zero-mass black
hole and the AdS solution, respectively \cite{16,17}. One expects
that this remarkable result remains true also in the generalized
supergravity \eq{3.1}.

\section{Concluding remarks} 

In this paper, we used the canonical formalism to investigate
asymptotic symmetries of $N=1+1$ sypersymmetric extension of 3D gravity
with torsion.

(1) We constructed the $N=1+1$ supergravity in a "non-Chern-Simons"
manner, starting directly from the topological Mielke-Baekler action
for 3D gravity with torsion. Consistency of the construction implies
various relations between coupling constants.

(2) It is a simple consequence of this analysis that the resulting
supergravity theory can be represented in terms of two independent
$Osp(1|2)$ CS theories. The corresponding gauge transformations are
rediscovered via the canonical approach in section V.

(3) We extended the AdS asymptotic conditions to the fermionic sector
and constructed the improved canonical generator. The resulting
asymptotic canonical algebra is expressed in terms of two independent
super-Virasoro algebras with different central charges.

Although $N=1+1$ extension of 3D gravity with torsion has a simple
structure, basic characteristics of its asymptotic symmetry can be
easily generalizad to $N=n+m$.

\section*{Acknowledgments} 

This work was supported by the Serbian Science Foundations. We would
like to thank A. Giacomini, R. Troncoso, S. Willison and A.
Mikovi\'c for useful comments regarding the reality conditions for
the $Osp(1|2)$ CS gauge theory.

\appendix
\section{Dirac matrices and spinors} 
\setcounter{equation}{0}

In 3D, Dirac matrices are two-dimensional. Since the tangent-frame
components of the metric are $\eta_{ij}=(+,-,-)$, all Dirac matrices
are chosen to be pure imaginary:
\bea
\g_0=\left(\ba{cc} 0 & i \\
                  -i & 0
           \ea\right)\, ,\qquad
\g_1=\left(\ba{cc} i & 0 \\
                   0 &-i
           \ea\right)\, ,\qquad
\g_2=\left(\ba{cc} 0 & i \\
                   i & 0
           \ea\right)\, .
\eea
They satisfy the relations
\bea
&&\{\g_i,\g_j\}=2\eta_{ij}\, ,\qquad
   [\g_i,\g_j]=2i\ve_{ij}{^k}\g_k\, ,                      \nn\\
&&\g_i^T=-\g_0\g_i\g_0\, ,\qquad \g_i^+=\g_0\g_i\g_0\, .   \nn
\eea

The Dirac adjoint of a spinor $\psi=(\psi_a)$ is
$\bar\psi=\psi^+\g_0$. Majorana spinors are defined by the condition
$\psi=C\bar\psi^T$, where $C$ is the charge conjugation matrix.
Spinors are taken to be real Majorana spinors ($C=-\g_0$), so
that $\bar\psi=\psi^T\g_0$. They satisfy the relations
$$
\bar\psi\chi=\bar\chi\psi\, ,\qquad
\bar\psi\g_i\chi=-\bar\chi\g_i\psi\, .
$$

Spinors are anticommuting objects (elements of a Grassmann algebra),
with an extension of the usual complex conjugation defined so that
$(\psi\chi)^*=\chi^*\psi^*$. Then, for Majorana spinors we have the
following reality conditions:
\be
(\bar\psi\chi)^*=\bar\psi\chi\, ,\qquad
(\bar\psi\g_m\chi)^*=-\bar\psi\g_m\chi\, .
\ee

The Fiertz identity has the form:
\be
(\bar\psi^1\g_i\psi^2)(\bar\psi^3\g^i\psi^4)
    = -2(\bar\psi^1\psi^4)(\bar\psi^2\bar\psi^3)
      -(\bar\psi^1\psi^2)(\bar\psi^3\psi^4)\,.
\ee

\section{Supersymmetric gauge transformations} 
\setcounter{equation}{0}

In this appendix, we discuss the form of $N=1+1$ super-Poincar\'e
gauge transformations and their commutator algebra.

\prg{1.} We start from the following SS gauge transformations on the
set of fields:
\bea
&&\d_0 b^i=iC\bar\ve\g^i\psi+iC'\bar\ve'\g^i\psi'\, ,      \nn\\
&&\d_0\om^i=iD\bar\ve\g^i\psi+iD'\bar\ve'\g^i\psi'\, ,     \nn\\
&&\d_0\psi^\Pi=4a(\nabla\ve^\Pi-imb^k\g_k\ve^\Pi)\, ,      \lab{B1}
\eea
where $C$, $C'$, $D$ and $D'$ are constant parameters. The action
\eq{2.4} is invariant under these transformations (up to surface
terms), provided the following relations are satisfied:
\bea
&&a(C-2g)+\a_3 D=0\, ,                                     \nn\\
&&\a_4 C+a(D-4gm)=0\, ,                                    \nn\\
&&-\L C-8agm^2+D\a_4=0\, ,                                 \nn
\eea
plus the corresponding ``primed" relations, obtained by replacing
$(C,D,g)$ with $(C',D',g')$. Using the simple normalization $C=1=C'$,
we find:
\bsubeq\lab{B2}
\bea
&&2ag=a+\a_3 D\, ,                                         \lab{B2a}\\
&&\a_4+a(D-2m)-2m\a_3D=0\, ,                               \nn\\
&&\L-2mDa+\a_4(D-2m)=0\, ,                                 \nn
\eea
and the additional three "primed" relations. Comparing the last 2
ordinary plus 2 ``primed" relations with the identities
$\a_4+ap+\a_3q=0$ and $\L+aq+\a_4p=0$, we obtain
\bea
&&D=-2m'\,,\qquad D'=-2m\, ,                               \nn\\
&&p=-2(m+m')\qquad q=4mm'\, .                              \lab{B2b}
\eea
\esubeq

For real and mutually different $m$ and $m'$, the effective
cosmological constant is negative:
$$
\L_{\rm eff}:=q-\frac{p^2}{4}=-(m-m')^2<0 \, .
$$
Without loss of generality, we can choose
\be
m-m'=\frac{1}{\ell}\, ,                                    \lab{B3}
\ee
which, in conjunction with \eq{B2a}, leads to \eq{3.4}.

The final form of the SS transformations is displayed in \eq{3.3}.

\prg{2.} An important information on the super-Poincar\'e gauge
transformations is obtained from their commutator algebra.

A direct calculation shows that the commutators $[\d_P,\d_P]$ and
$[\d_P,\d_S]$ close {\it off shell\/}, while the commutators
$[\d_S,\d_S]$ between two SS gauge transformations close only {\it on
shell\/}. Introducing a compact notation
$\d_{SP}(1)=\d_{SP}(\xi_1^\m,\th_1^i,\ve_1,\ve_1')$, and so on, and
denoting the sets of all fields and the field equations as in
\eq{3.7}, the commutator algebra of the super-Poincar\'e gauge
transformations takes the following form:
\bsubeq\lab{B4}
\be
\left[\d_{SP}(1),\d_{SP}(2)\right]\Phi^A_\m =\d_{SP}(3)\Phi^A_\m
 -4ia(\bar\ve_1\g^\r\ve_2+\bar\ve'_1\g^\r\ve'_2)E^A_{\m\r}\,,\lab{B4a}
\ee
where
\bea
\xi_3^\m&=&\xi_1\cdot\pd\xi_2^\m-\xi_2\cdot\pd\xi_1^\m
  +4ia(\bar\ve_1\g^\m\ve_2+\bar\ve_1'\g^\m\ve_2')\, ,       \nn\\
\th_3^i&=&\xi_1\cdot\pd\th_2^i-\xi_2\cdot\pd\th_1^i
   +\ve_{ijk}\th_1^j\th_2^k                                 \nn\\
 &&-4ia(\om^i{_\r}+2m'b^i{_\r})\bar\ve_1\g^\r\ve_2
   -4ia(\om^i{_\r}+2mb^i{_\r})\bar\ve'_1\g^\r\ve'_2         \nn\\
\ve_3&=&\xi_1\cdot\pd\ve_2-\xi_2\cdot\pd\ve_1
   +\frac{i}{2}(\th_1^m\g_m\ve_2-\th_2^m\g_m\ve_1)          \nn\\
 &&+i(\bar\ve_1\g^\r\ve_2+\bar\ve_1'\g^\r\ve_2')\psi_\r\, , \nn\\
\ve'_3&=&\ve_3\mbox{~~with~~}
   (\ve_1,\ve_2)\lra(\ve_1',\ve_2')\mbox{~~and~~}\psi\to\psi'\,.\lab{B4b}
\eea
\esubeq
Thus, the commutator algebra is closed {\it on shell\/}.

\section{Asymptotics of the fields} 
\setcounter{equation}{0}

Asymptotic conditions for a given field theory are not unique. Here,
we wish to illustrate how our asymptotics \eq{4.1} can be ``derived"
from the general principle \gp.

\prg{1.} In discussing the asymptotics of the gravitini fields, we
start with the on-shell equality
\bsubeq\lab{C1}
\be
\om^k+2m b^k\approx \tom^i+b^i{}/\ell
\ee
and the asymptotic formulas:
\bea
&&i\left(\tom^i{}_0+\frac{1}{\ell}b^i{}_0\right)\g_i
  =\frac{2r}{\ell^2}\s_- +\left(\ba{cc}
                                \cO_2 & \cO_1 \\
                                \cO_1 & \cO_2
                                \ea\right)\, ,             \nn\\
&&i\left(\tom^i{}_1+\frac{1}{\ell}b^i{}_1\right)\g_i
  =-\frac{1}{r}\s_3+\cO_3\, ,                              \nn\\
&&i\left(\tom^i{}_2+\frac{1}{\ell}b^i{}_2\right)\g_i
  =-\frac{2r}{\ell}\s_- +\left(\ba{cc}
                               \cO_2 & \cO_1 \\
                               \cO_1 & \cO_2
                               \ea\right)\, ,
\eea
\esubeq
where $i(\g_0-\g_2)=2\s_-$, $i\g_1=-\s_3$. Combining the form of the
leading-order terms in $\psi_\m$ with the relations $H_{\m\n}=\hcO$,
one obtains the following result:
\bea
&&\pd_0\Psi_2-\pd_2\Psi_0-\frac{r}{\ell^2}\s_-(\ell\Psi_0+\Psi_2)
   =\left(\ba{c}
          \cO_{5/2} \\
          \cO_{3/2}
          \ea\right)\, ,                                   \nn\\
&&\left(\ba{c}
            0 \\
    \ell^2 r^{-5/2}\pd_0\ups^-\ea\right)-\frac{r}{\ell^2}\s_-\Psi_1
   -\pd_1\Psi_0-\frac{1}{2r}\s_3\Psi_0=\cO_{7/2}\, ,       \nn\\
&&\left(\ba{c}
            0 \\
    \ell^2 r^{-5/2}\pd_2\ups^-\ea\right)-\frac{r}{\ell^2}\s_-\Psi_1
  -\pd_1\Psi_2-\frac{1}{2r}\s_3\Psi_2 = \cO_{7/2}\, .      \lab{C2}
\eea
As one can verify, this result is in agreement with the form of
higher-order terms in $\psi_\m$.

In a similar way we can verify the asymptotics for the ``primed''
gravitino field.

\prg{2.} If we substitute the asymptotic form of the gravitini fields
into $F^i=\hcO$ and $G^i=\hcO$, we obtain useful relations for the
bosonic fields. Since
$$
\bar\psi_0\g^i\psi_2=\cO_2\, ,\qquad
\bar\psi_1\g^i\psi_0=\cO_4=\bar\psi_1\g^i\psi_2\, ,
$$
and similarly for $\psi'_\m$, it turns out that the improved
asymptotic conditions for $b^i{_\m}$ and $\om^i{_\m}$ are of the same
form as in \cite{14}.

\section{Asymptotic parameters} 
\setcounter{equation}{0}

The asymptotic form of local parameters $(\xi^\m,\th^i,\ve,\ve')$ is
obtained by demanding that the super-Poincar\'e transformations
respect the asymptotic conditions \eq{4.1}. In other words, the
asymptotic parameters are defined by the following requirements:
\bea
&&\d_{SP}(\th^k,\xi^\r,\ve,\ve')b^i{_\m}
  =\d_0 B^i{_\m}\, ,                                       \nn\\
&&\d_{SP}(\th^k,\xi^\r,\ve,\ve')\om^i{_\m}
  =\d_0\Om^i{_\m}\, ,                                      \nn\\
&&\d_{SP}(\th^k,\xi^\r,\ve)\psi_\m
  =\d_0\langle\psi_\m\rangle+\d_0\Psi_\m\, ,               \nn\\
&&\d_{SP}(\th^k,\xi^\r,\ve')\psi'_\m
  =\d_0\langle\psi'_\m\rangle +\d_0\Psi'_\m\, .            \lab{D1}
\eea
Note that, according to \eq{4.1}, the leading-order terms in
$b^i{_\m}$ and $\om^i{_\m}$ are invariant, but those in $\psi_\m$ and
$\psi'_\m$ are not, as they contain non-invariant functions
$\chi^\mp$ and $v^\mp$.

Consider, first, the asymptotic form of the spinorial parameter
$\ve$. Combining the third relation in \eq{D1} with the adopted
asymptotics for $\psi$ and \eq{C1}, we conclude that $\ve$ has the
asymptotic form given in \eq{4.2}. Similar considerations determine
the asymptotic form of $\ve'$.

Real parts of the Poincar\'e parameters $\th^i,\xi^\m$ have been found
by analyzing the bosonic sector of the theory, hence, we can now
focus our attention on the terms in \eq{D1} that depend on Grassmann
odd variables and parameters. Without loss of generality, we can
assume that the souls of $\xi^\m$ and $\th^i$ \cite{20} do not appear
in the leading-order terms:
\bea
&&\xi^0_S=\cO_2=\xi^2_S\, ,\qquad \xi^1_S=\cO_1\, ,        \nn\\
&&\th^0_S=\cO_1=\th^2_S\, ,\qquad \th^1_S=\cO_2\, .        \lab{D2}
\eea
Now, the requirements \eq{D1}, applied to
$\d_0b^0{}_1,\d_0b^2{}_1,\d_0b^1{}_0,$ and $\d_0b^1{}_2$, read:
\bsubeq\lab{D3}
\bea
&&-\frac{\ell}{r}\th^2_S -\frac{r}{\ell}\pd_1\xi^0_S
  +\frac{i\ell^2}{r^2}(\eps^-\ups^-+\eps^+\ups^+)=\cO_4\, ,\nn\\
&&-\frac{\ell}{r}\th^0_S -r\pd_1\xi^2_S
  -\frac{i\ell^2}{r^2}(\eps^-\ups^--\eps^+\ups^+)=\cO_4\, ,\lab{D3a}\\
&&r\th^2_S+i\ell(\eps^-\chi^-+\eps^+\chi^+)=\cO_2\, ,      \nn\\
&&r\th^0_S+i\ell(\eps^-\chi^--\eps^+\chi^+)=\cO_2\, .      \lab{D3b}
\eea
Using the assumptions \eq{D2} in \eq{D3a}, we obtain:
\bea
&&2\xi^0_S=\frac{\ell^2}{r}\th^2_S+\frac{i\ell^3}{r^2}
  \left(\eps^-\ups^- -\eps^+\ups^+\right)+\cO_4\, ,        \nn\\
&&2\xi^2_S=\frac{\ell}{r}\th^0_S-\frac{i\ell^2}{r^2}
  \left(\eps^-\ups^- +\eps^+\ups^+\right)+\cO_4\, .        \lab{D3c}
\eea
\esubeq

The relations \eq{D3b} and \eq{D3c} are our final result for the souls
of $\xi^\m$ and $\th^i$. The remaining conditions in \eq{D1} do not
produce any further restrictions on the above result.

\section{The algebra of constraints} 
\setcounter{equation}{0}

The PB algebra of constraints is an important element of the
canonical struture of the theory. Having in mind that the part of the
algebra involving only bosonic constraints is of the same form as in
the bosonic theory \cite{14}, we skip that part here and display only
those (non-vanishing) PBs that involve at least one fermionic
constraint.

In the set of PBs involving only the primary constraints (see Table
1), the piece involving fermionic constraints reads:
\be
\{\bar f^\a,f^\b\}=-2g\ve^{0\a\b}\d\, , \qquad
\{\bar f'^\a,f'^\b \}=-2g'\ve^{0\a\b}\d\, .                 \lab{E1}
\ee
Looking at the set of PBs between the primary and secondary
constraints, we find that the subset involving fermionic constraints
is given by
\bea
&&\{\phi_i{}^\a,\cF\}=2igm\ve^{0\a\b}\g_i\psi_\b\d\, ,\qquad
  \{\Phi_i{}^\a,\cF\}=ig\ve^{0\a\b}\g_i\psi_\b\d\, ,       \nn\\
&&\{\bar f^\a,\cH_i\}=-2igm\ve^{0\a\b}\bar\psi_\b\g_i\d\, ,\qquad
  \{\bar f^\a,\cK_i\}=-ig\ve^{0\a\b}\bar\psi_\b\g_i\d\, ,  \nn\\
&&\{\bar f^\a ,\cF\}=-2g\ve^{0\a\b}\left[ -\pd_\b\d
  +\frac{1}{2}i(\om^n{}_\b+2mb^n{}_\b)\g_n\d\right]\, ,    \lab{E2}
\eea
plus the PBs obtained by the replacements $(\cF,\bar
f^\a)\to(\cF',\bar f'{}^\a)$, $(g,m)\to(g',m')$.

Finally, the part of the PB algebra involving only the first class
constraints ($\hat\cH_i,\hat\cK_i,\hat{\cF},\hat{\cF'}$) is given by
the relations (purely bosonic terms are omitted):
\bea
&&\{\hcH_i,\hcF\}=-im\g_i\hat{\cF}\d\, ,
  \qquad \{\hcH_i,\hcF'\}=-im'\g_i\hat{\cF'}\d\, ,         \nn\\
&&\{\hcK_i,\hcF\}=-\frac{i}{2}\g_i\hcF\d\, ,
  \qquad \{\hcK_i,\hcF'\}=-\frac{i}{2}\g_i\hcF'\d\, ,      \nn\\
&&\{\hcF_a,\hcF_b\}=
  -\frac{i}{4a}\,(\g_iC)_{ab}\left(\hcH_i-2m'\hcK_i\right)\d\,,\nn\\
&&\{\hcF'_a,\hcF'_b\}=
  -\frac{i}{4a}\,(\g_iC)_{ab}\left(\hcH_i-2m\hcK_i\right)\d\,.\lab{E3}
\eea

\end{document}